\documentclass{aa}
\usepackage{graphicx}
\begin{document}

\title{A Catalogue of infrared star clusters and stellar groups}

\author{E. Bica \inst{1}, C.M. Dutra \inst{2}, B. Barbuy \inst{2} }

\offprints{C.M. Dutra - dutra@astro.iag.usp.br}

\institute{Universidade Federal do Rio Grande do Sul, Instituto de F\'\i sica, CP\, 15051, Porto Alegre 91501-970, RS, Brazil\\
\mail{}
\and
 Universidade de S\~ao Paulo, Instituto de Astronomia, Geof\'\i sica e Ci\^encias atmosf\'ericas, CP\, 3386, S\~ao Paulo 01060-970, SP, Brazil\\
 \mail{}
 }

\date{Received 20/08/02; accepted 08/10/02}

\abstract{We compiled a catalogue of infrared  star clusters in the Galaxy, which are most of them embedded. It condenses the growing literature information. We also include in the sample infrared stellar groups which are less dense than star clusters, such as those embedded in the  dark clouds Taurus-Auriga and Chamaleon I. We provide galactic and equatorial coordinates,  angular dimensions, different designations and related objects such as nebulae. A total of 189 infrared clusters and 87 embedded stellar groups are included. A fraction of 25 \% of the embedded clusters are projected close to each other in pair or triplet systems, indicating that multiplicity plays an important role in star cluster formation. 
}

\titlerunning{Infrared star clusters and stellar groups}
\authorrunning{E. Bica et al.}

\maketitle

\section{Introduction}

The development of infrared detectors in the last decades has unveiled many optically
absorbed star clusters which are embedded in dark clouds or in more evolved star forming regions 
(Hodapp 1994). These clusters which are fully or partially immersed in interestellar gas 
and dust, are called embedded clusters (Lada \& Lada 1991). More generally infrared star clusters can be employed, since some of them can be older and not being embedded in complexes. Embedded clusters are ideal targets to investigate early stages of star formation and evolution. Many studies were made with newly identified 
clusters (e.g. Deharveng et al. 1999, Carpenter 2000 and Horner et al. 1997). It is 
important to gather all objects in a catalogue, which in turn will facilitate future 
developments, and avoid problems such as duplications. It allows the identification of less explored and 
unexplored areas that could be surveyed.
In particular, near infrared imaging surveys such as the Two Micron All Sky Survey (hereafter 2MASS, 
Skrutskie et al. 1997) and the Deep NIR Southern Sky Survey (DENIS; Epchtein et al. 1997) are 
providing material for cluster identifications (Soares \& Bica 2002, Reyl\'e \& Robin 2002). 

  Dutra \& Bica (2000a) identified 
58 new rather faint infrared star cluster candidates projected on the central 
parts of the Galaxy, using the 2MASS survey. The images employed  by 2MASS have a relatively low angular resolution and the fields near the
Galactic Centre are so crowded that is not possible to construct CMDs and
clearly establish the nature of these objects with the available material. 
Thus they are candidates to
infrared clusters to be verified with larger telescopes under good seeing,
and we preferred  not include them here. In a forthcoming study we
will present an infrared survey carried out at a higher angular resolution using the ESO NTT. More recently, Dutra \& Bica (2001) have made a search of objects 
in the 2MASS atlas in different parts of the sky finding 42 star clusters and stellar 
groups, which are included in the present study.

\begin{figure*} 
\resizebox{\hsize}{!}{\includegraphics{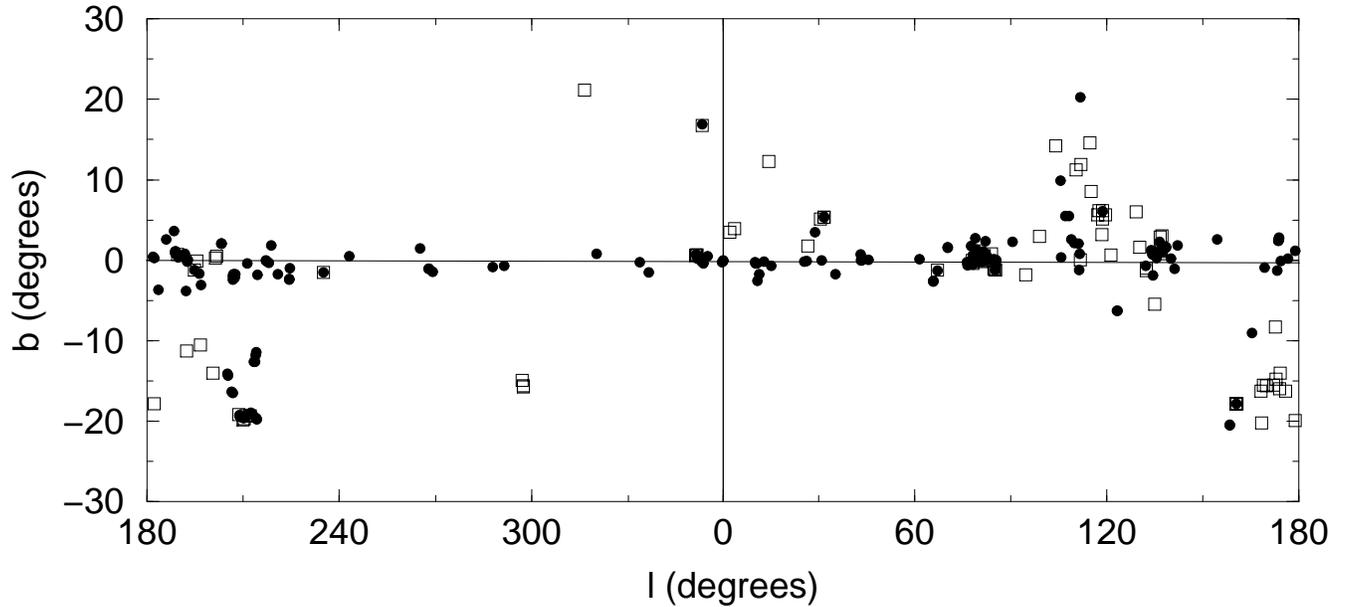}}
\caption[]{Angular distribution of infrared clusters (filled circles) and stellar groups (open squares). The Galactic plane and bulge minor axis direction are indicated by solid lines.}
\label{fig1}
\end{figure*}

Mermilliod (1996) compiled data on open clusters which are  updated in the Web interface {\em http://obswww.unige.ch/webda}. Recently, Mermilliod created a section for embedded clusters 
in his  star cluster database and included a few entries. He points out the need to compile 
data for these objects and  calls for contributions. One of the objectives of the present study 
is thus to compile a  catalogue of infrared star clusters.

Star cluster pairs and multiplets have been studied in detail in the LMC (Dieball et al. 2002), and  SMC (de Oliveira et al. 2000). Multiplicity  appears to 
be an important phenomenon which affects the dynamical evolution of young star clusters, e.g. 
in the SMC (de Oliveira et al. 2000). Some open cluster pairs appear to occur in the 
Galaxy (Subramaniam et al.  1995). However not much is known about the rate of cluster multiplicity at 
their birth, and infrared imaging of embedded clusters is now opening the possibility for such studies (e.g. the cluster pair NGC1333, Lada et al. 1996). Another objective of the present work is to verify cluster multiplicity in the infrared cluster catalogue.
   
In Sect. 2 we present the catalogue and its references. In Sect. 3 we discuss the properties of the sample such as angular and size 
distribution and cluster multiplicity. Finally, in Sect. 4 concluding remarks are given.

\section{The catalogue}

Infrared detectors and surveys have boosted the studies of infrared clusters, most of them embedded in dark or bright nebulae. It is important to gather these objects into a catalogue. During the search we realized that several objects were not as dense as star clusters, but appear to be physical systems. We illustrate the difference between the two object classes by means of a comparison of richness parameter for the DR22 Cluster and its  surrounding DR22 stellar group (Fig. 1 of  Dutra \& Bica 2001). The DR22 cluster is rather loose and uncrowded, an example of a less dense cluster.
For the cluster we obtained 54 stars (arcmin)$^{-2}$, while for the stellar group 25 stars (arcmin)$^{-2}$, as measured on the
2MASS atlas. No background or crowding correction was applied. Basically our classification in infrared star cluster or stellar group  follows a density estimation by eye on the available material. Some authors also adopted this classification  in  the object's original reference.
Other objects are concentrations of T Tauri stars (e.g. those in Taurus, Gomez et al. 1993 and Leinert et al. 1993, and Chamaleon, Whittet et al. 1991), also classified as groups. Consequently, the catalogue is formed by two object classes, the standard infrared star clusters and the less dense stellar groups. The present study includes 189 infrared star clusters and 87 infrared stellar groups from 86 references.

The nomenclature of infrared star clusters and groups is not uniform. Some authors adopted IRAS or AFGL related sources, dark nebulae, optical or radio HII regions, running numbers, among others. We included as many as possible designations. We preferred as first designation a related optical HII region, radio source, dark nebula and infrared sources, if multiple possibilities occur.

The catalogue has 276 objects and will be available as Table 1 in electronic form at CDS (Strasbourg) via anonymous ftp to {\rm cdsarc.u-strasbg.fr (130.79.128.5)}. Each entry occupies 2 catalogue lines. The first line shows by columns: (1) and (2) Galactic coordinates, (3) and (4) equatorial coordinates (J2000.0 epoch), (5) and (6) the major and minor axes dimensions in arcmin and  (7) object designations. On line 2 are included a class discriminator (IRC - infrared cluster or IRGr - infrared stellar group), comments and additional information such as distance in kpc (d=), age in Myr (t=), reddening (eFIR=) -- see below, linear diameter in pc (LD=) and a reference code (REFnnn). The references of the catalogue will be available as Table 2 in electronic form at CDS (Strasbourg) via anonymous ftp to {\rm cdsarc.u-strasbg.fr (130.79.128.5)}. It includes in Col. 1 the reference code from Table 1 and in Col. 2 the complete reference. Distances and ages were taken from the individual studies, when available. Angular diameters were measured on the images in the individual studies and on the 2MASS Atlas, when available. Reddening estimates based on CMDs are not systematically available. Consequently, we now  prefer  to provide  the  reddening estimates based on the object's coordinates extracted from the Schlegel et al.'s (1998) all-sky reddening maps, which are based on the 100$\mu$m dust emission. For each
cluster or stellar group we include the respective  E(B-V)$_{FIR}$ reddening value ("eFIR=" in Table 1)  associated 
to its  line of sight. We point out that this estimate is an
upper limit for the reddening, since it represents the total reddening in
the
object's line of sight, and background contributions may occur (Dutra \& Bica 2000b). We also point out that it  can be overestimated by
heated dust associated to  star forming regions (Dutra et al. 2002 and references therein).

\section{Sample Properties}

Overall properties of the catalogue can be inferred from the available data. We will analyse the angular distribution, distance and size distributions and multiplicity. Since the vast majority of the objects are embedded, the conclusions basically refer to the initial stages of star cluster and stellar group formation.

\subsection{Angular distribution}

Fig. 1 shows the angular distribution of the objects in galactic coordinates.  As expected infrared star clusters are projected close to the Galactic Plane. We note that the Anticentre zone and  Cygnus X ($\ell = 80^{\circ}$) are well explored. On the other hand, the southern Milky Way ($240^{\circ} <  \ell < 340^{\circ}$) requires a deep survey. The stellar groups tend to occur away from the Plane since they are mostly related to the nearby complexes (e.g. Orion at $\ell \approx$ 210$^{\circ}$ and $ b \approx$ -20$^{\circ}$,  Taurus  at $\ell \approx$ 175$^{\circ}$ and $ b \approx$ -15$^{\circ}$, and Chamaleon I at  $\ell \approx$ 297$^{\circ}$ and $ b \approx$ -15$^{\circ}$).

\subsection{Distance and Size distribution}

\begin{figure} 
\resizebox{\hsize}{!}{\includegraphics{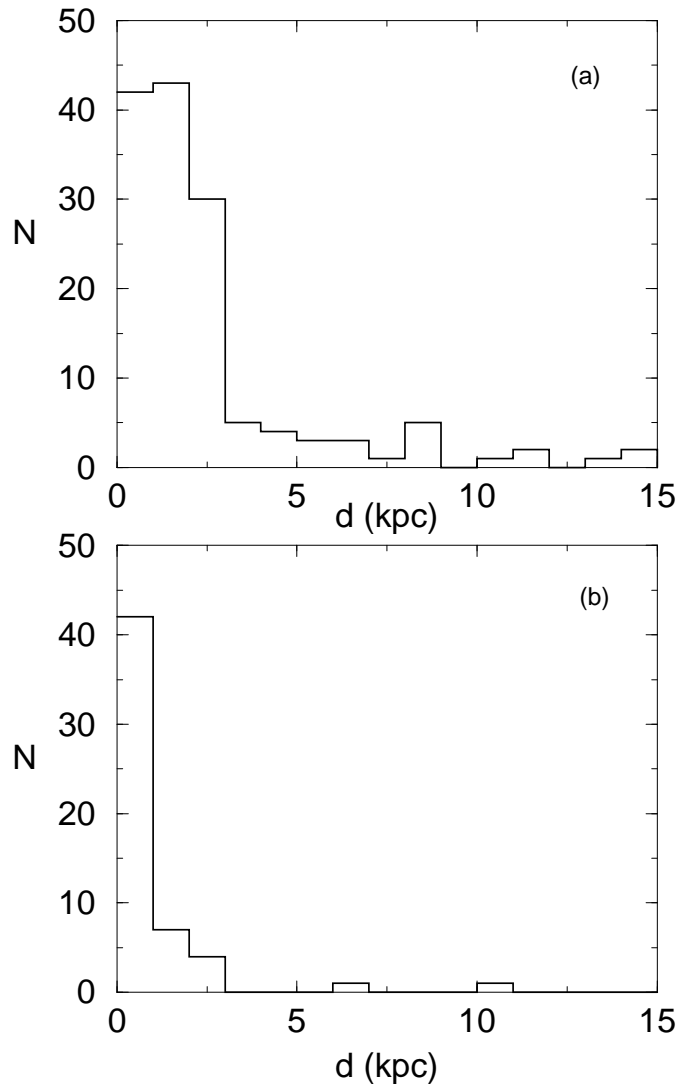}}
\caption[]{Distance histograms for: (a) infrared clusters and (b) stellar groups.}
\label{fig1}
\end{figure}

Fig. 2 shows the distance histograms for the infrared cluster and stellar group samples. Most objects in the cluster sample are located up to distances from the Sun of 3 kpc. The present sample is thus biased to relatively nearby zones. This shows the need to explore more distant disk regions. The histogram of stellar groups shows that they are mostly as far as 1 kpc,  since they are associated to nearby complexes (e.g. Taurus and Orion).

\begin{figure} 
\resizebox{\hsize}{!}{\includegraphics{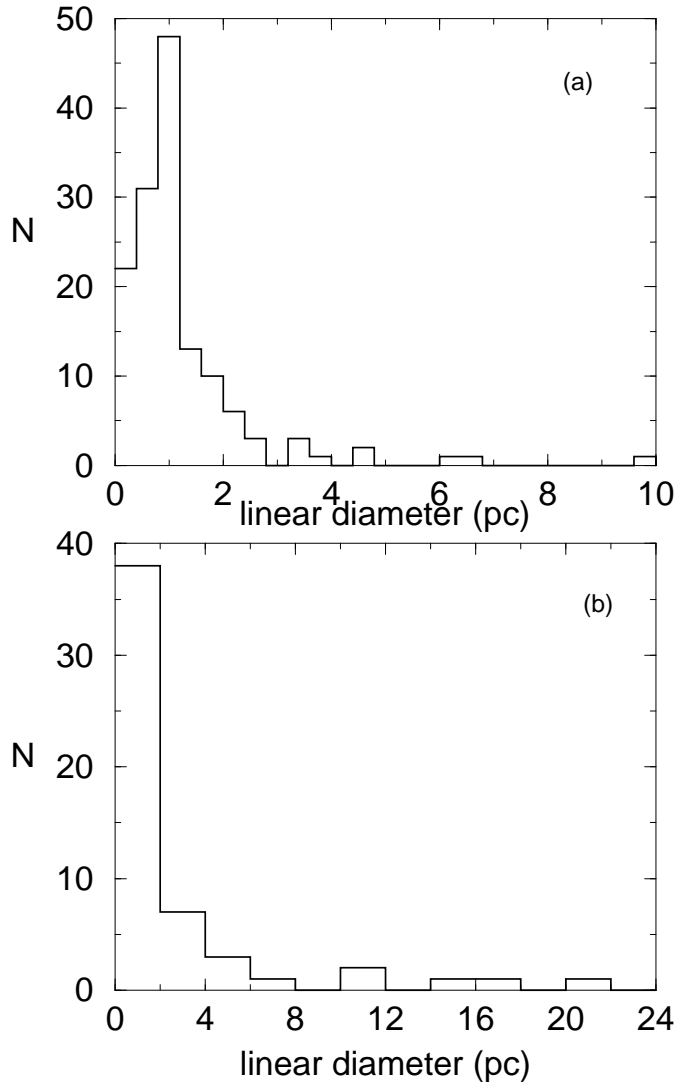}}
\caption[]{Linear diameter histograms for: (a) infrared clusters and (b) stellar groups.}
\label{fig1}
\end{figure}

We have used the distances and angular sizes provided in the catalogue to derive linear diameters for the clusters and stellar groups.
Fig. 3 shows the linear diameter histograms for the infrared clusters and stellar groups. The size distribution of clusters shows that the typical one is small. The peak diameter is $\approx$ 1 pc and the majority has linear diameter smaller than 3 pc. Open clusters older than 15 Myr are well distributed between 1 and 8 pc (Janes et al. 1988). Embedded clusters tend to be  smaller than evolved open clusters. 
This effect  appears to be physical since
a detection bias among the somewhat older open clusters is not expected for in this case the most
prominent IR clusters in a earlier stage would produce a series of prominent compact
open cluster population and this is not observed.  Nor a star formation bias is
expected since the Galaxy has no apparent reason to have currently formed
prominent clusters in excess. We are left with an age effect, related to the dynamical
evolution of the disk clusters in general.
Possibly we are witnessing an evolutionary effect from star cluster formation to those reflecting the internal dynamical evolution and the tidal effects of the Galaxy.

\subsection{Multiplicity}

During the present catalogue construction we realized that often embedded clusters have companions a few diameters away. These systems are indicated as pairs (mP) and triplets (mT) in Table 1. We found 18 pairs and 4 triplets resulting 48 clusters as components. Therefore, 25 \% the embedded cluster sample is formed in multiple systems. Not many open cluster pairs are known in the Galaxy (Subramaniam et al. 1995), consequently the dynamical evolution of double and triple embedded clusters probably leads to a high rate of mergers.

We also indicated groups of clusters and stellar groups which can have as many as 9 components (m9 in Table 1). Such ensembles can be related to the same molecular cloud such as the clusters in the Rosette Molecular Cloud (RMC), which has 8 cluster members. The separations in these groups tend to be larger then those in the pair and triplet systems.

The possible link between  duplicity effects and age and/or extinction selection
effects must await a larger number of
objects with CMD's and homogeneous age and  extinction determinations.
A comparison sample of non-embedded clusters, with ages of a
few 10$^7$ yr is also required.

\section{Conclusions}

We built a catalogue of infrared star clusters and stellar groups (less dense objects) in the Galaxy, using 86 references from the literature. A total of 276 entries are included, and we provide coordinates, angular dimensions, designations and additional information when available, such as distances and linear diameters.  We analysed overall properties of the sample. The angular distribution of the sample showed a deficiency of objects in the southern Milky Way indicating the need of a systematic search. The distance distribution indicated that the sample is biased to nearby zones ($ d < 3$ kpc). The Milky Way disk beyond this limit also requires more exploration. The linear size distribution revealed that embedded clusters are small, typically 1 pc. We noted that 25 \% of the embedded clusters occur in pairs or triplets with separation of a few diameters. Such proximity must affect their dynamical evolution, possibly resulting in mergers. For the first time a catalogue of this nature is presented and can be useful for future searches and developments.   

\begin{acknowledgements}
We acknowledge support from the Brazilian Institutions CNPq and FAPESP. CMD acknowledges FAPESP for a post-doc fellowship (proc. 00/11864-6).
\end{acknowledgements}

%
%sssssssssssssssssssssssssssss REFERENCESsssssssssssssssssssssssssssssss
%


\begin{thebibliography}{}

\bibitem[]{} Carpenter, J. M., 2000, AJ, 120, 3139
\bibitem[]{} Deharveng, L., Zavagno, A., Cruz-Gonz\'alez, I., et al. 1997, 317, 459
\bibitem[]{} de Oliveira, M.R., Dutra, C.M., Bica, E. \& Dottori, H. 2000, A\&AS, 146, 57
\bibitem[]{} Dieball, A., Mueller, H. \& Grebel, E. 2002, A\&A, 391, 547
\bibitem[]{} Dutra, C.M. \& Bica, E. 2000a, A\&A, 359, L9
\bibitem[]{} Dutra, C.M. \& Bica, E. 2000b, A\&A, 359, 347
\bibitem[]{} Dutra, C.M. \& Bica, E. 2001, A\&A, 376, 434
\bibitem[]{} Dutra, C.M., Santiago, B.X., Bica, E. \& Barbuy, B. 2002, MNRAS in press (astroph 0209168)
\bibitem[]{} Epchtein, N., de Batz, B., Capoani, L., et al. 1997, The Messenger, 87, 27
\bibitem[]{} Gomez, M., Hartmann, L., Kenyon, S.J. \& Hewett, R. 1993, AJ, 105, 1927
\bibitem[]{} Hodapp, K.-W. 1994, ApJS, 94, 615
\bibitem[]{} Horner, D.J., Lada, E.A. \& Lada, C.J. 1997, AJ, 113, 1788
\bibitem[]{} Janes, K.A., Tilley, C. \& Lyng\aa, G. 1988, AJ, 95, 771
\bibitem[]{} Lada C.J. \& Lada, E. A., 1991, in {\it ``The formation and evolution of star clusters'' }, 3
\bibitem[]{} Lada C.J., Alves, J. \& Lada, E. A. 1996, AJ, 111, 1964
\bibitem[]{} Leinert, Ch., Zinnecker, H., Weitzel, N., et al. 1993, A\&A, 278, 129
\bibitem[]{} Mermilliod J.C. 1996, in {\it `` The origins, evolution, and destinies of binary stars in clusters''}, ASP Conference Series, 90, 475
\bibitem[]{} Reyl\'e, C. \& Robin, A. C.  2002, A\&A, 384, 403
\bibitem[]{} Schlegel, D.J., Finkbeiner, D.P. \& Davis, M. 1998, ApJ, 500, 525
\bibitem[]{} Soares, J. B. \& Bica, E. , 2002, A\&A, 388, 172
\bibitem[]{} Subramaniam A., Gorti U. \& Sagar R. 1995, A\&A, 302, 86
\bibitem[]{} Skrutskie, M., Schneider, S.E., Stiening, R., et al. 1997, in {\it ``The Impact of Large Scale Near-IR Sky Surveys''}, ed. Garzon et al., Kluwer (Netherlands), 210, 187
\bibitem[]{} Whittet, C. B., Prusti, T. \& Wesselius, P. R., 1991, MNRAS, 249, 319


\end{thebibliography}
\end{document}